\documentclass[aps, prl, twocolumn, showpacs, superscriptaddress, floatfix]{revtex4}

\usepackage{graphicx}
\usepackage{dcolumn}
\usepackage{bm}
\usepackage{times}

\begin{document}

\title{Coupled Nd and B$'$ spin ordering in the double perovskites Nd$_2$NaB$'$O$_6$ (B$'$~$=$~Ru, Os)}

\author{A.A. Aczel}
\altaffiliation{author to whom correspondences should be addressed: E-mail:[aczelaa@ornl.gov]}
\affiliation{Quantum Condensed Matter Division, Neutron Sciences Directorate, Oak Ridge National Laboratory, Oak Ridge, TN 37831, USA}
\author{D.E. Bugaris}
\affiliation{Department of Chemistry and Biochemistry, University of South Carolina, Columbia, SC 29208, USA}
\author{J. Yeon}
\affiliation{Department of Chemistry and Biochemistry, University of South Carolina, Columbia, SC 29208, USA}
\author{C. de la Cruz}
\affiliation{Quantum Condensed Matter Division, Neutron Sciences Directorate, Oak Ridge National Laboratory, Oak Ridge, TN 37831, USA}
\author{H.-C. zur Loye}
\affiliation{Department of Chemistry and Biochemistry, University of South Carolina, Columbia, SC 29208, USA}
\author{S.E. Nagler}
\affiliation{Quantum Condensed Matter Division, Neutron Sciences Directorate, Oak Ridge National Laboratory, Oak Ridge, TN 37831, USA}
\affiliation{CIRE, University of Tennessee, Knoxville, TN 37996, USA}

\date{\today}

\begin{abstract}
We present a neutron powder diffraction study of the monoclinic double perovskite systems Nd$_2$NaB$'$O$_6$ (B$'$~$=$~Ru, Os), with magnetic atoms occupying both the A and B$'$ sites. Our measurements reveal coupled spin ordering between the Nd and B$'$ atoms with magnetic transition temperatures of 14~K for Nd$_2$NaRuO$_6$ and 16~K for Nd$_2$NaOsO$_6$. There is a Type I antiferromagnetic structure associated with the Ru and Os sublattices, with the ferromagnetic planes stacked along the c-axis and [110] direction respectively, while the Nd sublattices exhibit complex, canted antiferromagnetism with different spin arrangements in each system. 
\end{abstract}


\maketitle

\renewcommand{\topfraction}{0.85}
\renewcommand{\textfraction}{0.1}
\renewcommand{\floatpagefraction}{0.75}

Double perovskites with the formula A$_2$BB$'$O$_6$ have attracted considerable interest recently from both applied and fundamental physics perspectives. The excitement from the applied physics community has largely centered around the observation of room-temperature colossal magnetoresistance coupled with half-metallic behaviour in Sr$_2$FeMoO$_6$\cite{98_kobayashi}. These unique properties ensure a high degree of spin polarization of the charge carriers in a useful temperature regime applicable to potential spintronics devices\cite{98_prinz}. 

On the fundamental side, double perovskite systems provide excellent opportunities to study geometric frustration on the B$'$-site face-centered cubic (FCC) sublattice while tuning both the spin quantum number $S$ and spin-orbit coupling. Past efforts have found several exotic magnetic ground states, mostly in systems characterized by quantum spins of $S$~$=$~1/2 or 1. The diverse magnetism includes a collective singlet state coexisting with paramagnetism in Ba$_2$YMoO$_6$\cite{10_aharen, 11_carlo}, a collective singlet state in La$_2$LiReO$_6$\cite{10_aharen_2}, spin freezing without long-range order in Ba$_2$YReO$_6$\cite{10_aharen_2}, Sr$_2$MgReO$_6$\cite{03_wiebe} and Sr$_2$CaReO$_6$\cite{02_wiebe}, short-range order in La$_2$LiMoO$_6$\cite{10_aharen}, and a ferromagnetic (FM) Mott insulating state in Ba$_2$NaOsO$_6$\cite{02_stitzer, 07_erickson}. Theoretical studies have also indicated that a wealth of other magnetic ground states are possible in these 4$d$ and 5$d$ double perovskite quantum spin systems\cite{10_chen, 10_chen_2}.

\begin{figure}
\centering
\scalebox{0.17}{\includegraphics{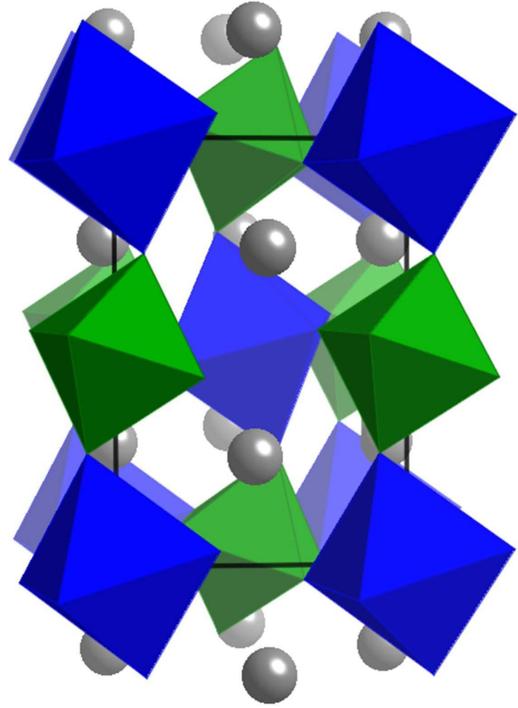}}
\caption{\label{Fig1} Double perovskite structure, with the large blue octahedra representing NaO$_6$, the small green octahedra depicting B$'$O$_6$, and the isolated gray spheres corresponding to Nd atoms. The small ionic radius of Nd leads to a large tilting of the NaO$_6$ and B$'$O$_6$ octahedra in Nd$_2$NaRuO$_6$ and Nd$_2$NaOsO$_6$.}
\end{figure} 

Some of the most interesting physical and magnetic properties observed in the double perovskite family can be attributed to the interactions between different types of magnetic atoms on the B and B$'$ sites. For example, the colossal magnetoresistance and half-metallic behaviour of Sr$_2$FeMoO$_6$ and related systems\cite{07_serrate} arises from an antiferromagnetic coupling between the delocalized 4$d$ or 5$d$ electrons and the localized 3$d$ electrons. Moreover, the systems Bi$_2$FeCrO$_6$\cite{05_baettig, 06_nechache, 12_bai} and Bi$_2$NiMnO$_6$\cite{05_azuma} are rare examples of multiferroics with both large spontaneous magnetization and polarization. Bi$_2$NiMnO$_6$ exhibits ferromagnetic properties, resulting from distributing two types of transition metal ions with and without $e_g$ electrons in a rock-salt configuration\cite{goodenough}, while Bi$_2$FeCrO$_6$ is a ferrimagnet. 


In contrast to the extensive studies performed on the subset of double perovskites with magnetic B and B$'$ sites, systems with magnetic ions on both the A and B$'$ sites remain largely unexplored territory. Exploratory synthesis via solid state reactions and hydroxide flux methods\cite{12_bugaris} have produced a series of these materials, including R$_2$NaRuO$_6$\cite{04_gemmill}, R$_2$NaOsO$_6$\cite{05_gemmill}, R$_2$LiOsO$_6$\cite{06_gemmill} and R$_2$MgIrO$_6$\cite{10_mugavero} (R~$=$~rare earth), but the magnetic properties have generally only been investigated with bulk probes such as magnetic susceptibility. One exception is the recent work of Ref.~\cite{09_makowski}, where neutron diffraction was used to investigate magnetic ordering in the systems R$_2$LiRuO$_6$ (R~$=$~Nd, Tb, Pr). The authors found that all compositions showed evidence for long-range magnetic order except for the Gd system, with the Ru sublattice exhibiting common Type I antiferromagnetic (AF) order and the spins of the rare earth atoms forming a canted arrangement. It is also interesting to note that the magnetic transitions involved coupled spin ordering of the two magnetic sublattices with $T_c$'s~$>$~10~K, while in several double perovskite systems with only an A-site rare earth magnetic sublattice, no magnetic ordering has been observed in susceptibility measurements down to 2 K\cite{04_davis, 05_mugavero}. These observations indicate that the magnetic behaviour of the rare earth ions in these systems is very sensitive to the presence of other types of magnetic atoms.   

The syntheses, crystal structures (shown in Fig.~\ref{Fig1}) and magnetic susceptibilities of the monoclinic double perovskites Nd$_2$NaRuO$_6$ and Nd$_2$NaOsO$_6$ were recently reported\cite{04_gemmill, 05_gemmill}. These materials combine a small A site cation with a relatively large ionic radii difference between the B and B$'$ sites, and therefore exhibit very large, room-temperature structural distortions relative to most other double perovskites. Comparing the Nd$_2$NaRuO$_6$ and Nd$_2$NaOsO$_6$ magnetization and magnetic susceptibility results to La$_2$NaRuO$_6$ and La$_2$NaOsO$_6$ clearly shows that the addition of the magnetic rare earth sublattice dramatically alters the magnetic properties of these materials. For example, the small deviation from the Curie-Weiss law with decreasing temperature in the susceptibility of La$_2$NaRuO$_6$ is replaced by a large ferromagnetic-like increase in the susceptibility for Nd$_2$NaRuO$_6$, and significant magnetic hystersis is observed in the M vs. H curve for the Nd system that is absent in the La case. Furthermore, a ferromagnetic-like increase with decreasing temperature in the susceptibility of La$_2$NaOsO$_6$ gives way to complicated magnetic behaviour for Nd$_2$NaOsO$_6$, likely indicative of a spin-flop transition for an applied field of 1.7~T at T~$=$~2~K. 

In this work, neutron powder diffraction (NPD) measurements have been performed on the double perovskites Nd$_2$NaRuO$_6$ and Nd$_2$NaOsO$_6$. These specific materials were chosen in an effort to better understand how the unconventional magnetism in La$_2$NaRuO$_6$ and La$_2$NaOsO$_6$\cite{13_aczel} evolves when an exchange path between the A and B$'$ sites is added to these materials. To produce a large enough yield of both materials for the diffraction experiments, solid state synthesis methods were required. For polycrystalline Nd$_2$NaRuO$_6$, Nd$_2$O$_3$ (Alfa Aesar, 99.99\%) was first activated by heating in air at 1000$^\circ$C for 12 h, Na$_2$CO$_3$ (Mallinckrodt, 99.95\%) was dried overnight at 150$^\circ$C, and RuO$_2$ was prepared by heating Ru (Engelhard, 99.95\%) in air at 1000$^\circ$C for 24 h. The starting materials were then mixed together in a 1:0.55:1 ratio, and this included a 10\% molar excess of Na$_2$CO$_3$ to offset the volatilization of Na$_2$O during heating. This mixture was heated to 500$^\circ$C in 1 h, held at 500$^\circ$C for 8 h, heated to 900$^\circ$C in 1 h, and held at 900$^\circ$C for 12 h before turning off the furnace and allowing the sample to cool to room temperature. Powder X-ray diffraction showed the presence of Nd$_2$NaRuO$_6$, as well as Nd$_2$O$_3$ and Nd$_3$RuO$_7$ impurities. Therefore, the sample was ground together with additional Na$_2$CO$_3$ and RuO$_2$, and then subjected to the same heating profile.  Due to remaining impurities as confirmed by powder X-ray diffraction, this process was repeated twice more.  Finally, all the peaks in the powder pattern could be indexed in the monoclinic space group {\it P2$_1$/n} with lattice parameters in good agreement with those reported for Nd$_2$NaRuO$_6$\cite{04_gemmill}.  

\begin{figure}
\centering
\scalebox{0.45}{\includegraphics{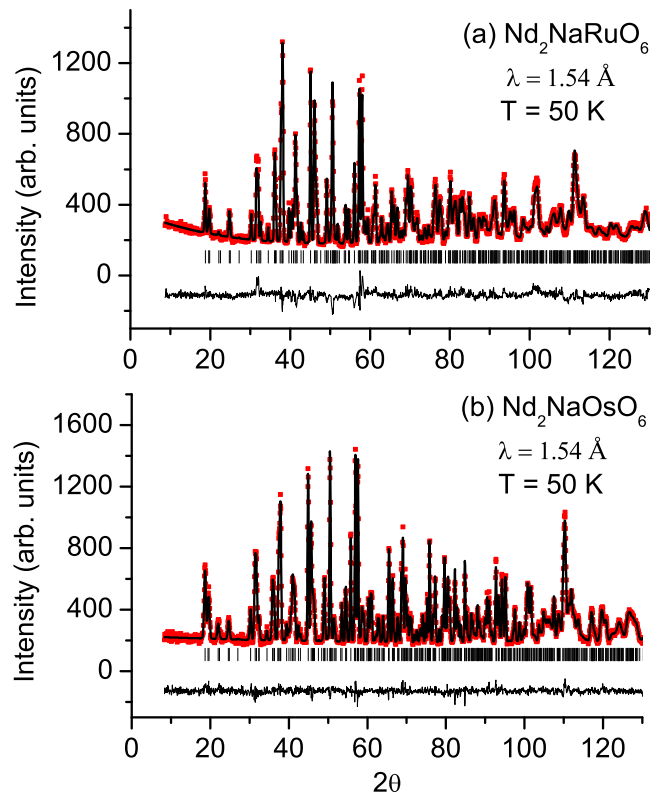}}
\caption{\label{Fig2} Neutron powder diffraction measurements with a wavelength of 1.54~\AA~at T~$=$~50~K for both (a) Nd$_2$NaRuO$_6$ and (b) Nd$_2$NaOsO$_6$. }
\end{figure}

For polycrystalline Nd$_2$NaOsO$_6$, the starting materials were nearly identical, with the only difference being that Os (J\&J Materials Inc.) replaced RuO$_2$. The heat treatment was also slightly modified, as the starting mixture was heated to 700$^\circ$C in 1 h and held at 700$^\circ$C for 6 h, before turning off the furnace and allowing the sample to cool to room temperature. Powder X-ray diffraction showed the presence of Nd$_2$NaOsO$_6$, as well as Nd$_2$O$_3$ and Nd$_3$OsO$_7$ impurities. Therefore, the sample was ground together with additional Na$_2$CO$_3$ and Os, and then subjected to the same heating profile. Due to remaining impurities as confirmed by powder X-ray diffraction, this process was repeated four more times.  Finally, all the peaks in the powder pattern could be indexed in the monoclinic space group {\it P2$_1$/n} with lattice parameters in good agreement with those reported for Nd$_2$NaOsO$_6$\cite{05_gemmill}.  

For the neutron powder diffraction experiment, roughly 5 g of each polycrystalline sample was loaded in a closed-cycle refrigerator and studied using the HB-2A powder diffractometer at the High Flux Isotope Reactor of Oak Ridge National Laboratory. Data from HB-2A were collected with neutron wavelengths $\lambda$~$=$~1.54~\AA~and $\lambda$~$=$~2.41~\AA~at temperatures of 4 - 50 K using a collimation of 12$'$-open-6$'$. The shorter wavelength gives a greater intensity and higher $Q$ coverage that was used to investigate the crystal structures in this low temperature regime, while the longer wavelength gives lower $Q$ coverage and greater resolution that was important for investigating the magnetic structures of these materials. The NPD data was analyzed using the Rietveld refinement program FullProf\cite{93_rodriguez} and the representational analysis software SARAh\cite{00_wills}. 

Our neutron diffraction results for Nd$_2$NaRuO$_6$ and Nd$_2$NaOsO$_6$ reveal that the Nd and B$'$ spins order at the same temperature with very similar magnetic structures to the Nd$_2$LiRuO$_6$ analogue\cite{09_makowski}. More specifically, the B$'$ sublattices are characterized by Type I antiferromagnetic order with the ferromagnetic planes stacked along the c-axis and the [110] direction in the Ru and Os compositions respectively, while the Nd sublattices form different canted arrangements in the two materials. The extra Nd-O-B$'$ exchange interactions in these materials have a significant effect on the magnetism, effectively destroying the delicate balance of extended superexchange interactions that lead to the incommensurate magnetic structure of La$_2$NaRuO$_6$ and the drastically-reduced ordered moment of La$_2$NaOsO$_6$.  

Figure~\ref{Fig2} shows $\lambda$~$=$~1.54~\AA~NPD data at T~$=$~50~K for monoclinic Nd$_2$NaRuO$_6$ and Nd$_2$NaOsO$_6$, while Table I shows the refinement parameters extracted from the T~$=$~4~K data. Comparing these results with previous X-ray diffraction data collected on single crystals at room temperature\cite{04_gemmill, 05_gemmill} reveals no evidence for a structural phase transition between 300 and 4~K in either material, and the structural distortion remains large at all temperatures as indicated by the 4~K $\beta$ values of 90.898(3)$^\circ$ and 90.870(1)$^\circ$ for the Ru and Os systems respectively. The Rietveld refinements confirm that there is essentially no site mixing between the Na and B$'$ atomic positions, as expected for double perovskite systems with a charge difference of +4 between the B and B$'$ sites\cite{93_andersen}. There was also no site mixing found between the Nd and Na atomic positions.

\begin{centering}
\begin{table}[htb]
\caption{Structural parameters for Nd$_2$NaRuO$_6$ and Nd$_2$NaOsO$_6$ at T~$=$~4~K extracted from the $\lambda$~$=$~1.54~\AA~neutron powder diffraction data. } 

\medskip

(a) Nd$_2$NaRuO$_6$ \\
Space group {\it P2$_1$/n} \\ 
a~$=$~5.4808(2)~\AA \\
b~$=$~5.9016(2)~\AA \\ 
c~$=$~7.8728(3)~\AA \\ 
$\beta$~$=$~90.898(3)$^\circ$ \\
$\chi^2$~$=$~6.12 \\
R$_{wp}$~$=$~8.31~\% \\

\begin{tabular}{| c | c | c | c | c |}
\hline 
Atom & Site & x & y & z  \\  \hline
Nd & 4e & 0.4837(8) & 0.0750(5) & 0.2534(6)    \\  
Na & 2a & 0 & 0 & 0  \\  
Ru & 2b & 0.5 & 0.5 & 0  \\  
O$_1$ & 4e & 0.209(1) & 0.328(1) & 0.0518(7) \\  
O$_2$ & 4e & 0.614(1) & 0.4512(9) & 0.2298(7) \\  
O$_3$ & 4e & 0.331(1) & 0.781(1) & 0.0660(7) \\ \hline  
\end{tabular}

\medskip

(b) Nd$_2$NaOsO$_6$ \\
Space group {\it P2$_1$/n} \\ 
a~$=$~5.5128(1)~\AA \\
b~$=$~5.8932(1)~\AA \\ 
c~$=$~7.9242(1)~\AA \\ 
$\beta$~$=$~90.870(1)$^\circ$ \\
$\chi^2$~$=$~1.14 \\
R$_{wp}$~$=$~5.50~\% \\

\begin{tabular}{| c | c | c | c | c |}
\hline 
Atom & Site & x & y & z  \\  \hline
Nd & 4e & 0.4818(4) & 0.0698(3) & 0.2533(3)    \\  
Na & 2a & 0 & 0 & 0  \\  
Os & 2b & 0.5 & 0.5 & 0  \\  
O$_1$ & 4e & 0.2086(5) & 0.3268(5) & 0.0504(3) \\  
O$_2$ & 4e & 0.6155(5) & 0.4498(4) & 0.2307(3) \\  
O$_3$ & 4e & 0.3380(5) & 0.7805(4) & 0.0658(3) \\  \hline
\end{tabular}
\end{table}
\end{centering}

\begin{figure*}
\centering
\scalebox{0.88}{\includegraphics{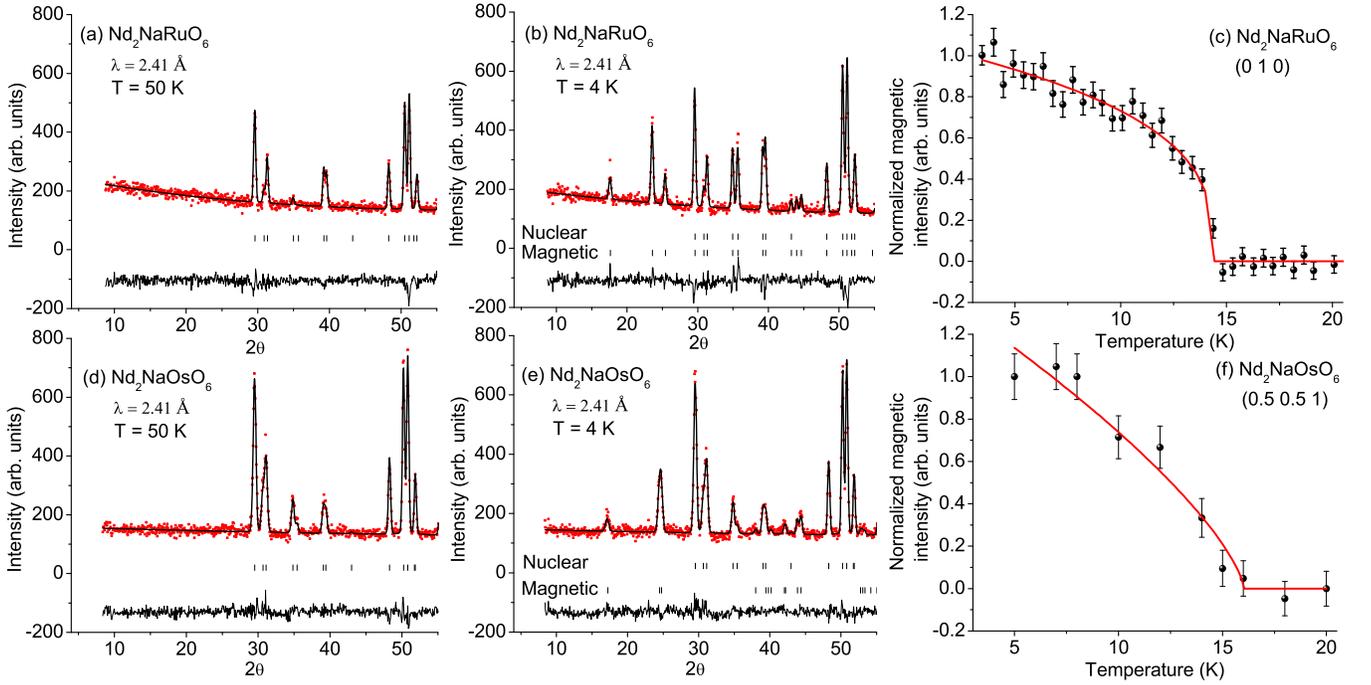}}
\caption{\label{Fig3} Neutron powder diffraction data with $\lambda$~$=$~2.41~\AA~is shown for Nd$_2$NaRuO$_6$ at (a) T~$=$~50 K and (b) T~$=$~4~K. The magnetic refinement at 4~K corresponds to the $\Gamma_1$ representation. (c) A plot of the normalized magnetic intensity at the (0 1 0) position for Nd$_2$NaRuO$_6$, showing an increase in intensity below the magnetic transition at T~$=$~14~K. Neutron powder diffraction data with $\lambda$~$=$~2.41~\AA~is shown for Nd$_2$NaOsO$_6$ at (d) T~$=$~50 K and (e) T~$=$~4~K. The magnetic refinement at 4~K corresponds to the $\Gamma_2$ - $\Gamma_4$ model, as explained in the text. (f) A plot of the normalized magnetic intensity at the (0.5 0.5 1) position for Nd$_2$NaOsO$_6$, showing an increase in intensity below the magnetic transition at T~$=$~16~K. The solid curves in (c) and (f) are guides to the eye.}
\end{figure*}

As indicated by the $\lambda$~$=$~2.41~\AA~NPD data presented in Fig.~\ref{Fig3}, at low temperatures additional scattering is observed in both the Nd$_2$NaRuO$_6$ and Nd$_2$NaOsO$_6$ neutron diffraction patterns at commensurate positions. The peaks can be indexed on the basis of a propagation vector $\vec{k}$~$=$~0 for the Ru system and $\vec{k}$~$=$~$(0.5~0.5~0)$ for the Os analogue. Fig.~\ref{Fig3}(c) shows the temperature-dependence of the normalized magnetic intensity for the Nd$_2$NaRuO$_6$ (0 1 0) magnetic reflection, revealing a magnetic transition temperature of $\sim$~14~K. A similar plot of the (0.5 0.5 1) magnetic reflection for Nd$_2$NaOsO$_6$, shown in Fig.~\ref{Fig3}(f), indicates a magnetic transition temperature of $\sim$~16~K. For the Os material, a series of powder diffraction patterns were collected between 4 and 20~K to investigate the potential change in the magnetic structure suggested by previous susceptibility measurements around 10~K\cite{05_gemmill}. However, only a smooth increase in the intensities of the magnetic peaks with decreasing temperature was observed.

\begin{figure*}
\centering
\scalebox{0.25}{\includegraphics{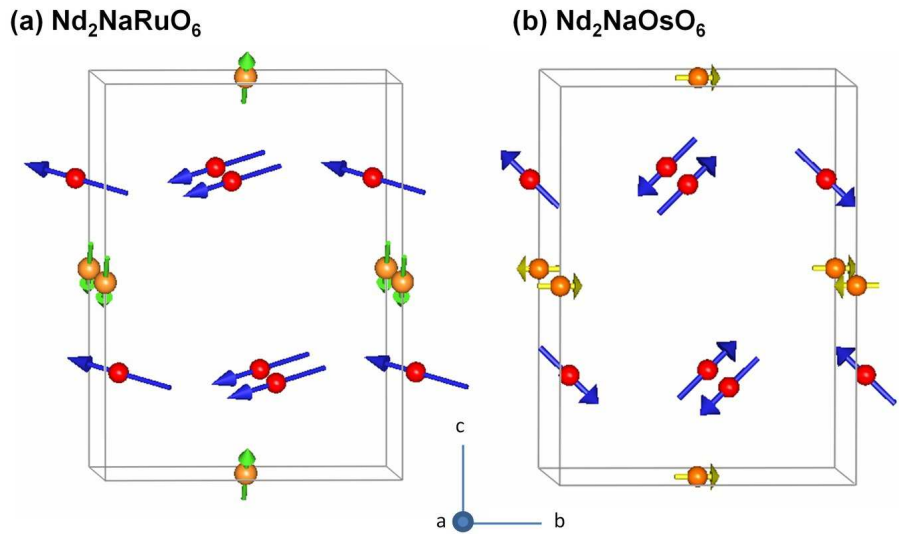}}
\caption{\label{Fig4} The magnetic structures determined from the neutron powder diffraction data for (a) Nd$_2$NaRuO$_6$ and (b) Nd$_2$NaOsO$_6$ ($\Gamma_2$ - $\Gamma_4$ model). The B$'$ sublattices exhibit Type I AF order, while the Nd spins form different canted arrangements. In each picture, the longer arrows represent the Nd moments and the shorter arrows represent the B$'$ moments. }
\end{figure*}

Representational analysis allowed the possible magnetic structures for these materials to be constrained on the basis of the crystal symmetry. The diffraction data was first modeled assuming magnetic ordering of only the Ru or Nd moments at the transition, but it quickly became clear that these assumptions could not explain the data. There are only two irreducible representations consistent with coupled Nd and Ru spin ordering and the observed propagation vector for Nd$_2$NaRuO$_6$; these are $\Gamma_1$ and $\Gamma_3$ in Kovalev's notation\cite{kovalev}. One of the main differences between these two magnetic structures is the spin components that are coupled ferromagnetically. More specifically, the spin components along the b-axis are ferromagnetically-coupled for both sublattices in the $\Gamma_1$ representation, while the spin components along both the a- and c-axes are ferromagnetically-coupled in the $\Gamma_3$ representation. The neutron diffraction data shows no magnetic intensity at the (020) position, while there is significant magnetic intensity at the (200) and (002) positions. These observations are only consistent with the $\Gamma_1$ spin model, and this refinement is illustrated in Fig.~\ref{Fig3}(b) with the corresponding magnetic structure shown in Fig.~\ref{Fig4}(a). The Nd spins form a canted arrangement with a moment size of 2.31(3)~$\mu_B$ per Nd. The best refinement results in a Type I AF ground state for the Ru sublattice, with an ordered Ru moment of 1.62(8)~$\mu_B$. Since the Ru$^{5+}$ magnetic form factor has not been accurately measured, the estimate for $<j_0>$ given in Ref.~\cite{03_parkinson} was used for this refinement. The data was also refined using the Ru$^{3+}$ and Ru$^{4+}$ magnetic form factors from Ref.~\cite{LosAlamos_table} and the Ru moment was found to be 1.72(8)~$\mu_B$ and 1.67(8)~$\mu_B$ in these two cases. These results indicate that the refined Ru ordered moment is relatively insensitive to the specific Ru magnetic form factor used in the analysis. The magnetic properties of the Ru sublattice are very similar to the magnetism reported for several other Ru$^{5+}$ double perovskites\cite{89_battle, 83_battle, 84_battle, 00_izumiyama, 00_doi}, and the coupled Ru-Nd spin ordering is exactly what was observed in the related double perovskites R$_2$LiRuO$_6$\cite{09_makowski}.

\begin{centering}
\begin{table}[htb]
\caption{Magnetic models and refined magnetic moments for Nd$_2$NaRuO$_6$ and Nd$_2$NaOsO$_6$. } 

\begin{tabular}{| c | c | c | c | }
\hline 
 & Ru $\Gamma_1$ &  Os $\Gamma_2$ + $\Gamma_4$ & Os $\Gamma_2$ - $\Gamma_4$  \\  \hline
B$_1'$ & [v$_1$, v$_2$, v$_3$] & [v$_1$, v$_2$, v$_3$] & [v$_1$, v$_2$, v$_3$]    \\  
B$_2'$ & [-v$_1$, v$_2$, -v$_3$] & [-v$_1$, v$_2$, -v$_3$] & [v$_1$, -v$_2$, v$_3$]  \\  
Nd$_1$ & [v$_1$, v$_2$, v$_3$] & [v$_1$, v$_2$, v$_3$] & [v$_1$, v$_2$, v$_3$]  \\  
Nd$_2$ & [-v$_1$, v$_2$, -v$_3$] & [v$_1$, -v$_2$, v$_3$] & [-v$_1$, v$_2$, -v$_3$]  \\  
Nd$_3$ & [v$_1$, v$_2$, v$_3$] & [v$_1$, v$_2$, v$_3$] & [v$_1$, v$_2$, v$_3$] \\  
Nd$_4$ & [-v$_1$, v$_2$, -v$_3$] & [v$_1$, -v$_2$, v$_3$] & [-v$_1$, v$_2$, -v$_3$] \\ \hline 
B$_1'$, $\mu_x$ & 1.29(6) & 0 & 0  \\
$\mu_y$ & 0 & 0 & 0.9(1) \\
$\mu_z$ & 1.0(1) & -0.64(9) & 0  \\
$\mu$ & 1.62(8) & 0.64(9) & 0.9(1) \\ \hline
Nd$_1$, $\mu_x$ & -1.05(4) & 0.1(1) & 0.2(2)  \\
$\mu_y$  & -1.98(3) & 1.20(6) & 1.16(9)  \\
$\mu_z$  & 0.56(5) & -1.18(8) & -1.1(1) \\
$\mu$ & 2.31(3) & 1.69(7) & 1.6(1) \\ \hline
R$_{mag} (\%) $ & 10.8 & 18.1 & 16.8 \\ \hline
\end{tabular}

\end{table}
\end{centering}



A similar situation was encountered for the Os sample, as no suitable magnetic models were found when assuming that only the Nd or Os spins ordered at the magnetic transition. There are also only two magnetic models allowed by symmetry that are consistent with coupled Nd and Os spin ordering and the observed propagation vector for Nd$_2$NaOsO$_6$. In Kovalev's notation\cite{kovalev}, these two models are superpositions of the $\Gamma_2$ and $\Gamma_4$ irreducible representations and can be described as $\Gamma_2$ + $\Gamma_4$ and $\Gamma_2$ - $\Gamma_4$. The basis vectors and the refined moments for these two models are indicated in Table II. A non-zero Os moment is critical to account for the full intensity of the magnetic peak at (0.5 0.5 0) in each case. The magnetic refinements were performed using the Os$^{5+}$ magnetic form factor from Ref.~\cite{11_kobayashi}. The $\Gamma_2$ - $\Gamma_4$ model gives the best refinement, with the result illustrated in Fig.~\ref{Fig3}(e) and a picture of the magnetic structure depicted in Fig.~\ref{Fig4}(b). In this model, the Nd and Os moments refine to 1.6(1)~$\mu_B$ and 0.9(1)~$\mu_B$ respectively, with the Os sublattice forming a Type I AF arrangement with the spins aligned along the b-axis. The Nd moment for this system is significantly reduced compared to Nd$_2$NaRuO$_6$, suggesting that the size of the ordered B$'$ moment affects the magnitude of the rare earth moment. Unlike Nd$_2$NaRuO$_6$, the stacking of the ferromagnetic planes associated with the Type I AF order is along the [110] direction.  


The B$'$ sublattices of Nd$_2$NaRuO$_6$ and Nd$_2$NaOsO$_6$ exhibit conventional Type I AF order for S~$=$~3/2 B$'$ double perovskites despite the large monoclinic structural distortions. This is in sharp contrast to the unconventional magnetism found in the highly-distorted compounds La$_2$NaRuO$_6$ and La$_2$NaOsO$_6$\cite{13_aczel}, characterized by incommensurate magnetism and a drastically-reduced ordered moment respectively. The Nd compounds actually have even larger monoclinic structural distortions than their La counterparts, but the B$'$ magnetic ground states are exactly what one expects for undistorted systems with dominant B$'$-O-O-B$'$ AF exchange interactions. The addition of the rare earth magnetic sublattice seems to destroy the delicate balance of competing exchange interactions giving rise to the unconventional magnetism in the La systems and instead stabilizes simple, commensurate AF on the B$'$ sublattice. The rare earth ions introduce Nd-O-B$'$ interactions into the systems. Although these new couplings are expected to be weaker than B$'$-O-O-B$'$ extended superexchange, and this is supported by the nearly identical magnetic transition temperatures between the La$_2$NaB$'$O$_6$ and Nd$_2$NaB$'$O$_6$ compounds, they are strong enough to act as a significant perturbation and alter the magnetic ground state of the B$'$ sublattice. The resulting magnetic structures are very similar to the spin arrangements reported for R$_2$LiRuO$_6$\cite{09_makowski}. 

Both the ordered Ru and Os moments of Nd$_2$NaRuO$_6$ and Nd$_2$NaOsO$_6$ are significantly reduced compared to the expected spin only value of 3~$\mu_B$. Due to the orbital singlet ground states of the B$'$ atoms of these systems, spin-orbit coupling should have a negligible effect on reducing the ordered moment size. However, strong covalency effects are often associated with 4$d$ and 5$d$ magnetic systems and have been suggested to lead to smaller moments in these materials, particularly in iridates\cite{09_kim} and osmates\cite{12_calder, 12_calder_2}. In fact, the Os atoms of Nd$_2$NaOsO$_6$ and NaOsO$_3$\cite{12_calder} are found in nearly identical, slightly-distorted oxygen octahedral cages so covalency effects should be very similar in the two systems. This is consistent with the comparable Os ordered moments of the two materials: 1.0(1)~$\mu_B$ for NaOsO$_3$ and 0.9(1)~$\mu_B$ for Nd$_2$NaOsO$_6$. Furthermore, the difference in the B$'$ ordered moment size for Nd$_2$NaOsO$_6$ and Nd$_2$NaRuO$_6$ can also be explained by covalency effects, as these are expected to be enhanced in the osmate, relative to the ruthenate, due to the extended nature of the Os 5$d$ orbitals.  

In conclusion, we have investigated the magnetic properties of the double perovskites Nd$_2$NaRuO$_6$ and Nd$_2$NaOsO$_6$ with neutron powder diffraction. Despite the large monoclinic structural distortions of these materials, the B$'$ sublattices exhibit conventional Type I AF order, as observed for several S~$=$~3/2 B$'$ magnetic double perovskites. The Nd and B$'$ ions are also found to order at the same magnetic transition temperature. The magnetic Nd sublattice creates Nd-O-B$'$ couplings, effectively destroying the delicate balance of extended superexchange interactions that lead to unconventional magnetism in the highly-distorted double perovskites La$_2$NaRuO$_6$ and La$_2$NaOsO$_6$. These results indicate that A-site magnetic rare earths play an essential role in the determination of the magnetic ground state for the B$'$ sublattice in double perovskite systems. 

\begin{acknowledgments}
This research was supported by the US Department of Energy, Office of Basic Energy Sciences. A.A.A., C.d.l.C. and  S.E.N. were supported by the Scientific User Facilities Division. The neutron experiments were performed at the High Flux Isotope Reactor, which is sponsored by the Scientific User Facilities Division. D.E.B., J.Y. and H.z.L. would like to acknowledge financial support through the Heterogeneous Functional Materials for Energy Systems (HeteroFoaM) Energy Frontiers Research Center (EFRC), funded by the US Department of Energy, Office of Basic Energy Sciences under award number DE-SC0001061.
\end{acknowledgments}

\end{document}